\documentclass[aps,preprint,preprintnumbers,prb,amssymb,floatfix]{revtex4-1}
\usepackage{graphicx}
\usepackage{dcolumn}
\usepackage{bm}
\usepackage{amsmath}
\usepackage{amsfonts}
\usepackage{amssymb}
\usepackage{ulem}
\usepackage{color}
\usepackage{caption}
\usepackage{subfig}
\usepackage{hyperref}
\usepackage{wasysym}

\begin{document}

\captionsetup[subfigure]{labelformat=empty}

\title{Thermal and electronic transport characteristics of highly stretchable graphene kirigami}
\author{Bohayra Mortazavi$^{(a)}$}
\email[Corresponding author: ]{bohayra.mortazavi@gmail.com}
\affiliation{Institute of Structural Mechanics, Bauhaus-Universit\"{a}t Weimar, Marienstr. 15, D-99423 Weimar, Germany}
\author{Aur\'{e}lien Lherbier$^{(a)}$}
\email[Corresponding author: ]{aurelien.lherbier@uclouvain.be}
\affiliation{Universit\'e catholique de Louvain, Institute of Condensed Matter and Nanosciences, Chemin des \'etoiles 8, 1348 Louvain-la-Neuve, Belgium}
\author{Zheyong Fan$^{(a)}$}
\affiliation{School of Mathematics and Physics, Bohai University, Jinzhou, China}
\affiliation{COMP Centre of Excellence, Department of Applied Physics, Aalto University, Helsinki, Finland}
\author{Ari Harju}
\affiliation{COMP Centre of Excellence, Department of Applied Physics, Aalto University, Helsinki, Finland}
\author{Timon Rabczuk}
\affiliation{College of Civil Engineering, Department of Geotechnical Engineering, 
Tongji University, Shanghai, China}
\author{Jean-Christophe Charlier}
\affiliation{Universit\'e catholique de Louvain, Institute of Condensed Matter and Nanosciences , Chemin des \'etoiles 8, 1348 Louvain-la-Neuve, Belgium}

\date{\today ; $(a)$: These three first co-authors contributed equally.}

\begin{abstract}
For centuries, cutting and folding the papers with special patterns 
have been used to build beautiful, flexible and complex three-dimensional structures. 
Inspired by the old idea of kirigami (paper cutting), 
and the outstanding properties of graphene, 
recently graphene kirigami structures were fabricated to enhance the stretchability of graphene. 
However, the possibility of further tuning the electronic and thermal transport 
along the 2D kirigami structures have remained original to investigate. 
We therefore performed extensive atomistic simulations to explore the electronic, 
heat and load transfer along various graphene kirigami structures. 
The mechanical response and thermal transport were explored 
using classical molecular dynamics simulations. 
We then used a real-space Kubo-Greenwood formalism 
to investigate the charge transport characteristics in graphene kirigami. 
Our results reveal that graphene kirigami structures 
present highly anisotropic thermal and electrical transport. 
Interestingly, we show the possibility of tuning the thermal conductivity 
of graphene by four orders of magnitude. 
Moreover, we discuss the engineering of kirigami patterns 
to further enhance their stretchability by more than 10 times 
as compared with pristine graphene. 
Our study not only provides a general understanding concerning 
the engineering of electronic, thermal and mechanical response of graphene 
but more importantly can be useful to guide future studies 
with respect to the synthesis of other 2D material kirigami structures, 
to reach highly flexible and stretchable nanostructures 
with finely tunable electronic and thermal properties.
\end{abstract}


\maketitle

\section {Introduction}\label{section_intro}
The two-dimensional form of sp$^2$ carbon atoms, 
so called graphene ~\cite{Novoselov_Science_2004,Geim_Nature_2007,CastroNeto_RMP_2009},
is widely considered as a wonder material owing to its exceptional mechanical ~\cite{Lee_Science_2008}, 
thermal ~\cite{Balandin_NatMater_2011} and electronic ~\cite{CastroNeto_RMP_2009} properties. 
Graphene in its single-layer and free-standing form exhibits 
a unique combination of ultra-high mechanical and thermal conduction properties, 
which include a Young's modulus of about 1000 GPa, 
intrinsic strength of about 130 GPa ~\cite{Lee_Science_2008} 
and thermal conductivity of around 4000 W/mK ~\cite{Gosh_NatMater_2010}, 
outperforming all-known materials. 
These unique properties of graphene offer it as a promising candidate for 
a wide-variety of applications such as simultaneously enhancing the thermal, 
electronic and mechanical properties of polymer nanocomposites, 
stretchable nanoelectronics, nanosensors, nanoresonators 
and nanoelectromechanical systems (NEMS). 
The great achievements by graphene, promoted the successful synthesis of 
other high-quality 2D materials like hexagonal boron-nitride ~\cite{Kubota_Science_2007}, 
silicene ~\cite{Aufray_APL_2010,Vogt_PRL_2012}, 
phosphorene ~\cite{Das_ACSNano_2014,Li_NatNano_2014}, 
borophene ~\cite{Mannix_Science_2015}, 
germanene ~\cite{Bianco_ACSNano_2013}, 
and transition metal dichalcogenides (TMDs) like MoS$_{2}$ ~\cite{Radisavljevic_NatNano_2011}. 
An interesting fact about graphene lies in its unique ability 
to present largely tunable and in some cases contrasting properties through 
mechanical straining ~\cite{Guinea_SSC_2012,Metzger_NL_2010,Pereira_PRL_2009,BarrazaLopez_SSC_2013,Guinea_NatPhys_2010}, 
defect engineering ~\cite{Banhart_ACSNano_2011,Lherbier_PRB_2012,Cummings_AdvMater_2014,Cresti_NanoResearch_2008} 
or chemical doping ~\cite{Wehling_NL_2008,Lherbier_PRL_2008,Wang_NL_2008,Schedin_NatMater_2007,Miao_NL_2012,Soriano_2DMat_2015}.\\

Apart from the latest scientific advances during the last decade, 
Origami is an old Chinese and Japanese art with a simple idea of 
transforming a flat paper into a complex and elaborated three-dimensional structure 
through folding and sculpting techniques. 
This idea is well-explained in the Japanese word of Origami, 
which consists of ``ori'' meaning ``folding'', and ``kami'' meaning ``paper''. 
Kirigami (``kiru'' means cutting) is another variation of origami 
which includes solely the cutting of the paper, 
and therefore that is different than origami 
which is achieved only by paper folding. 
As mentioned earlier, graphene presents 
exceptionally high mechanical strength and stiffness, 
but the stretchability of graphene is limited because of 
its brittle failure mechanism ~\cite{Lee_Science_2008,Zhang_NatComm_2014}. 
For many applications such as those related to 
flexible nanoelectronics, the building blocks are requested to 
present ductile mechanical response and 
the brittle nature of graphene acts as a negative factor. 
Therefore, engineering of the graphene structure to enhance 
its ductility and more generally to provide tunable mechanical response 
can provide more opportunities for graphene applications. 
To address this issue and inspired by the ancient idea of kirigami, 
an exciting experimental advance has been recently achieved 
with respect to the fabrication of the graphene kirigami, 
through employing the lithography technique ~\cite{Blees_Nature_2015}. 
Experimental results confirmed the considerable enhancement 
in the stretchability and bending stiffness of graphene 
through applying the kirigami approach ~\cite{Blees_Nature_2015}. 
This experimental advance consequently raised the importance of 
theoretical investigations to provide more in-depth physical insight. 
Using classical molecular dynamics simulations, 
both mechanical ~\cite{Qi_PRB_2014} and thermal conductivities ~\cite{Wei_Carbon_2016} 
of graphene kirigami were studied. In these investigations ~\cite{Qi_PRB_2014,Wei_Carbon_2016}, 
likely to the experimental set-up ~\cite{Blees_Nature_2015}, 
graphene films with only few linear cuts were investigated. 
However, in another amazing experimental study ~\cite{Shyu_NatMater_2015}, 
kirigami approach with periodic and curved-cuts 
were employed for the engineering of elasticity in polymer nanocomposites. 
This recent experimental work on the polymer nanocomposites kirigami ~\cite{Shyu_NatMater_2015} 
films consequently highlights the possibility of fabrication of 
graphene kirigami structures with networks of curved cuts 
rather than few straight cuts (as it was proposed in the earlier investigation ~\cite{Blees_Nature_2015}). 
Because of the complexities of such experimental set-ups, 
theoretical studies can be considered as promising alternatives 
to shed light on the properties of these structures. 
To the best of our knowledge, mechanical, thermal and electronic transport properties 
in graphene kirigami nanomembranes with periodic and 
curved cutting patterns have not been studied. 
This study therefore aims to explore the transport properties 
in the graphene kirigami with different cutting patterns using
large scale atomistic simulations. 

\section {Models and Methods}\label{section_methods}

\subsection{Graphene kirigami structural design}
\begin{figure}[ht]
\includegraphics[width=\columnwidth]{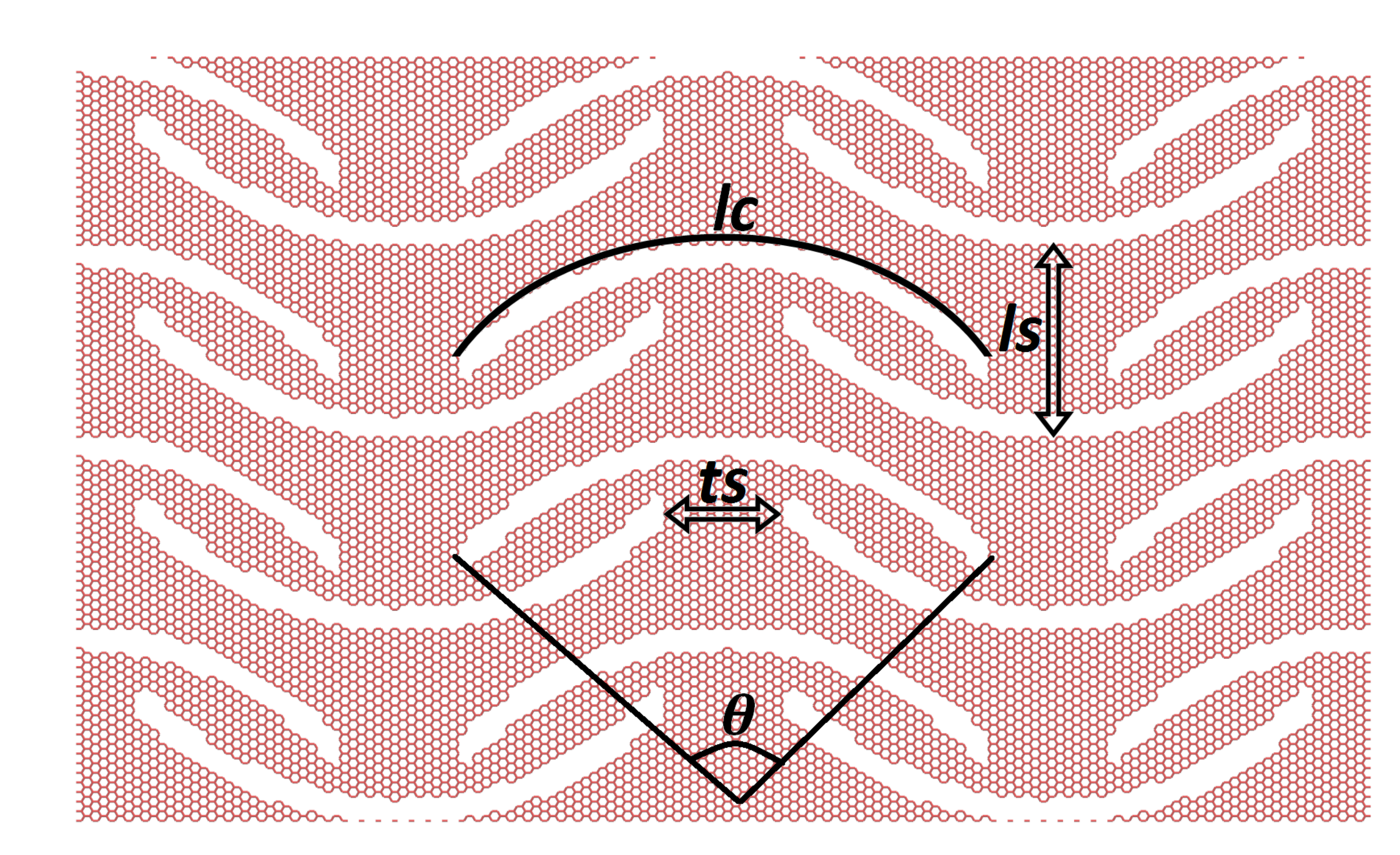}
\caption{Atomistic and periodic structure of graphene kirigami, 
to illustrate the key geometric parameters: the curvature angle ($\theta$), 
the longitudinal and transverse spacing distances ($l_s$ and $t_s$), respectively.
}
\label{fig1}
\end{figure}

The graphene kirigami structures explored in this study 
were constructed by considering periodic boundary conditions 
such that they represent graphene films with repeating cuts ~\cite{Shyu_NatMater_2015}. 
In order to describe our modeling strategy, in Fig.\ref{fig1}, 
a sample of constructed molecular model of graphene kirigami is illustrated, 
along with the geometric parameters. 
The length of the cuts ($l_c$) is the first parameter that describes the graphene kirigami. 
The width of the cuts is accordingly changed 
to adjust the volume fraction of the removed materials form the graphene. 
In this work, we considered curved cuts which are defined by 
the curvature angle ($\theta$ in Fig.\ref{fig1}). 
This way, curvature angles of 0 and 180 degrees 
represent straight and half-circular cuts, respectively. 
The longitudinal ($l_s$ in Fig.\ref{fig1}) and transverse ($t_s$ in Fig.\ref{fig1}) 
spacing distances between two adjacent cuts are two other 
key parameters used to characterize the graphene kirigami structures. 
In our modeling, the cut length was considered as the main factor, 
which directly determines the system scale. 
Therefore by changing the cut length we scaled the other parameters 
of kirigami to keep the structural pattern comparable.

\subsection{Stretchability and thermal transport properties}
We employed classical molecular dynamics (MD) simulations 
to evaluate mechanical and thermal transport properties of 
graphene kirigami structures. 
The atomic interaction between carbon atoms was modelled 
using the Tersoff potential ~\cite{Tersoff_PRB_1988,Tersoff_PRL_1988} 
optimized by Lindsay and Broido ~\cite{Lindsay_PRB_2010}. 
Among different possibilities for the modeling of graphene structures, 
the optimized Tersoff potential is highly efficient and can be used to simulate large scale systems. 
More importantly, it gives reasonable predictions of both thermal conductivity and 
mechanical properties of pristine graphene as compared with experimental 
results ~\cite{Xu_NatComm_2014,Mortazavi_Carbon_2016,Fan_PRB_2015}. 
In our classical MD simulations, we applied 
periodic boundary conditions along the x-y planar directions. 
The thickness of graphene kirigami nanomembranes was taken as 3.35 \AA.
We employed the equilibrium molecular dynamics (EMD) method 
to calculate the thermal conductivity of the graphene kirigami structures, using GPUMD ~\cite{Fan_arXiv_2016},
an efficient MD code fully-implemented on graphics processing units,
to compute the thermal conductivity of large scale graphene kirigami structures. 
We calculated the heat flux vector with the appropriate form for many-body potentials
as described in ~\cite{Fan_PRB_2015}:
\begin{equation}
\mathbf{J}= \sum_{i}\sum_{j\neq i}\left(\mathbf{r}_j-\mathbf{r}_i\right)\left(\frac{\partial U_j}{\partial \mathbf{r}_{ji}} \cdot \mathbf{v}_i\right),
\end{equation}
where $\mathbf{v}_i$ is the velocity of atom $i$, 
$\mathbf{r}_i$ is the position vector of atom $i$, 
$\mathbf{r}_{ji}$ is the position vector from atoms $j$ to $i$, 
and $U_j$ is the potential energy associated with atom $j$. 
The thermal conductivity tensor can be acquired based on 
the Green-Kubo formula:
\begin{equation}
\kappa_{\alpha\beta} = \frac{1}{V k_B T^2} \int_0^{\infty} \langle \mathbf{J}_{\alpha}(0)\mathbf{J}_{\beta}(t) \rangle dt \label{GKformula},
\end{equation}
where $k_B$ is Boltzmann's constant, 
$T$ is the simulation temperature, 
and $V$ is the total volume of the system.
Thermal transport in graphene kirigami nanomembranes is anisotropic, 
and the thermal conductivities for a given sample 
were calculated in both the longitudinal and transverse directions. 
We note that when using the EMD method, 
accurately predicting the converged thermal conductivity requires doing many 
independent simulations and averaging them~\cite{Schelling_PRB_2002}. 
We remind that because of the use of an inaccurate heat-flux formula
for many-body interactions, the LAMMPS~\cite{Plimpton_JCompPhys_1995} implementation of the 
EMD method significantly underestimates the thermal conductivity of graphene ~\cite{Fan_PRB_2015}.\\

Then, we evaluated the mechanical response of graphene kirigami structures 
by performing uniaxial tensile tests at room temperatures. 
In this case we used LAMMPS~\cite{Plimpton_JCompPhys_1995}, a free and open-source package. 
As it was discussed in a recent study \cite{Mortazavi_Carbon_2016}, 
we modified the inner cutoff of the Tersoff potential from 0.18 nm to 0.20 nm, 
which was found to accurately reproduce the experimental results 
for the mechanical properties of pristine graphene. 
The time increment of these simulations was set as 0.2 fs. 
Before applying the loading conditions, all structures were relaxed 
and equilibrated using the Nos\'{e}-Hoover barostat and thermostat method (NPT ensemble). 
For the loading condition, we increased the periodic size of 
the simulation box along the loading direction at every time step 
by an engineering strain rate of $5\times10^8$ s$^{-1}$. 
Simultaneously, in order to ensure the uniaxial stress condition, 
the stress along the transverse direction was controlled 
using the NPT ensemble ~\cite{Mortazavi_Carbon_2016}. 
To report the stress-strain relations, 
we calculated the Virial stresses every 2 fs 
and then averaged them over every 20 ps intervals. 

\subsection{Electronic transport properties}
The electronic and transport calculations, 
including electronic densities of states (DOS) per unit of surface 
($\rho(E)=\text{Tr}[\delta( E -\hat{H})]/S$) 
and Kubo-Greenwood conductivities ($\sigma(E,L)$), rely on a 
well established real-space Kubo-Greenwood method ~\cite{Mayou_JPhysIFrance_1995,
Triozon_PRB_2002,Latil_PRL_2004,Leconte_PRB_2011,Lherbier_PRB_2012,Uppstu_PRB_2014}.
The electronic framework is based on a tight-binding (TB) Hamiltonian described by 
a $\pi$-$\pi^*$ TB orthogonal model (one orbital per carbon atom) with 
interactions up to the third nearest neighbors (hopping terms are obtained
as $\gamma(d)=\gamma_0\,\,e^{-3.37\,\left(\frac{d}{d_{CC}}-1\right)}$,
with $\gamma_0=-2.8$ eV and $d_{CC}=1.42$ \AA~) ~\cite{Pereira_PRB_2009,Lherbier_NR_2013}.
The Kubo-Greenwood approach combines a Lanczos recursion scheme 
to obtain the spectral quantities as the DOS~\cite{Haydock_CompPhysComm_1980} 
and a Chebyshev polynomial expansion method to compute the 
time-dependent electronic diffusivity ($D(t)$) 
through the evaluation of the time evolution operator 
$\hat{U}(t) = \prod_{n=0}^{N-1}\text{exp}(i\hat{H}\Delta t/\hbar)$,
with $\Delta t$ the chosen time step. 
The total diffusivity is expressed as $D=D_{\text{x}}+D_{\text{y}}$ with 
$D_{\text{x}}(E,t)=\frac{\partial \Delta X^2(E,t)}{\partial t}$ and 
$\Delta X^2(E,t) = \text{Tr}[\delta (E-\hat{H})
\vert \hat{X}(t)-\hat{X}(0)\vert^2 ]/\text{Tr}[\delta (E-\hat{H})]$.
Traces are evaluated by averaging on eight random phase wave packets ($N_{\text{rpwp}}=8$) on large enough systems.
More explicitely, for a given operator $\hat{A}$ one has 
$\text{Tr}[\hat{A}] = \sum_{i=1}^{N_{\text{orb}}}\langle \phi_i \vert \hat{A} \vert \phi_i \rangle 
\backsimeq \frac{N_{\text{orb}}}{N_{\text{rpwp}}} \sum_{j=1}^{N_{\text{rpwp}}} \langle \varphi_j \vert \hat{A} \vert \varphi_j \rangle $,
where $\vert \phi_i \rangle$ is the $i^{\text{th}}$ orbital over $N_{\text{orb}}$ orbitals 
and $\vert \varphi_j \rangle=\frac{1}{\sqrt{N_{\text{orb}}}}\sum_{i=1}^{N_{\text{orb}}}\vert \phi_i \rangle e^{i\theta_{i}^{r}}$ 
is a random phase wave packet ($\theta_{i}^{r}$ is a random angle).
At short time, $D_{\text{x,y}}(t)$ is linear in time 
with the slope being the average square velocity v$_{\text{x,y}}^2(t=0)$.
In the following, the considered graphene kirigami structures 
are constructed by repeating periodically a rectangular unit cell
whose dimensions increases with the cut length ($l_c=$ 10, 20, 40, 80, and 160 nm).
Because of this perfect periodicity, i.e. without the introduction
of any stochastic disorder, the long time behavior of the diffusivity $D(t)$
should be in the ballistic regime, i.e. a linear increase in time 
with the slope being v$_{\text{x,y}}^2(t=\infty)$.
These two ballistic regimes and its associated velocities 
(v$^2(t=0)$ and v$^2(t=\infty)$) are in principle different.
The short time velocity corresponds to the dynamics of charge at a small scale,
i.e. graphene honeycomb lattice, while the long time velocity corresponds
to the propagation associated to Bloch states of the periodic kirigami structure.
However, to observe this second ballistic signature, the simulated
time propagation has to be very long, especially for the large cuts,
such that the periodicity effects at large scale become sufficiently significant.
Actually, because of computing limitations this second ballistic regime is
not always reached. In between these two ballistic regimes 
an intermediate regime should be observed and may include (pseudo) diffusive and/or localization regimes,
that is a (slow increase) saturation of $D(t)$ and/or a decrease of $D(t)$, respectively.
These regimes correspond to the fact that at intermediate length scale,
the electron wave propagation encounter scattering related to the presence
of cuts and interferences related to these multiple scattering events.
Similar transient phenomena was observed in a previous study on carbon nanotubes 
doped periodically with nitrogen atoms ~\cite{Khalfoun_PRB_2015}.
The determination of mean elastic scattering times and paths
($\tau$ and $l_e$, respectively) is not easily practicable in 
systems containing two periodic patterns, which are in the present case 
the honeycomb and the kirigami lattices,
because of the complex dynamics of charge carriers. 
Instead of characterizing the systems with the evaluation of $\tau$ and $l_e$,
it is preferable to describe the transport with a well-defined quantity which is
the Kubo-Greenwood length-dependent conductivity ($\sigma(E,L)$).
Since the kirigami structures present anisotropy, it is also 
important to consider independently the two longitudinal components
of the conductivity tensor ($\sigma_{\text{xx}}(E,L)$ and $\sigma_{\text{yy}}(E,L)$).
The former reads $\sigma_{\text{xx}}(E,L) = 
\frac{1}{2}e^2\rho(E)\frac{\partial \Delta X^2(E,t)}{\partial t}\vert_{t=t_L \,\,\text{with}\,\,\, L=2\sqrt{\Delta X^2(E,t_L)}}$.

\section{Results and discussions}

\subsection{Thermal transport}

We first study heat transport along the graphene kirigami films. 
Since in the EMD method periodic boundary conditions are applied 
and the systems are at equilibrium  
(no external heat-flux or temperature boundary conditions are applied), 
this method is much less sensitive to the finite-size effects 
as compared to the non-equilibrium molecular dynamics (NEMD) method \cite{Schelling_PRB_2002}
or the approach-to-equilibrium method (AEMD) \cite{Lampin_JApplPhys_2013}. 
As discussed in a recent study \cite{Fan_arXiv2_2016}, 
the EMD method predicts a thermal conductivity of 2900$\pm$100 W/mK 
for pristine graphene at room temperature. 
In this case, a relatively small simulation cell size (with $\sim$ 10$^4$ atoms) 
is sufficient to eliminate the finite-size effects \cite{Fan_PRB_2015}. 
However, using the NEMD or the AEMD method, the thermal conductivity of graphene 
keeps increasing by increasing the sheet length 
and the thermal conductivity was found to not fully converge even up to 0.1 mm \cite{Barbarino_PRB_2015}. 
Interestingly, for the graphene kirigami structures 
with periodic cutting patterns, the EMD method can be used 
to evaluate the full thermal conductivity tensor by constructing a single sample of moderate size. 
Whereas using the NEMD or AEMD method the length dependency of 
the predicted thermal conductivity values should be assessed, 
which demands high computational costs associated with 
the modeling of several large samples, each with different lengths.
\begin{figure}[ht]
\includegraphics[width=\columnwidth]{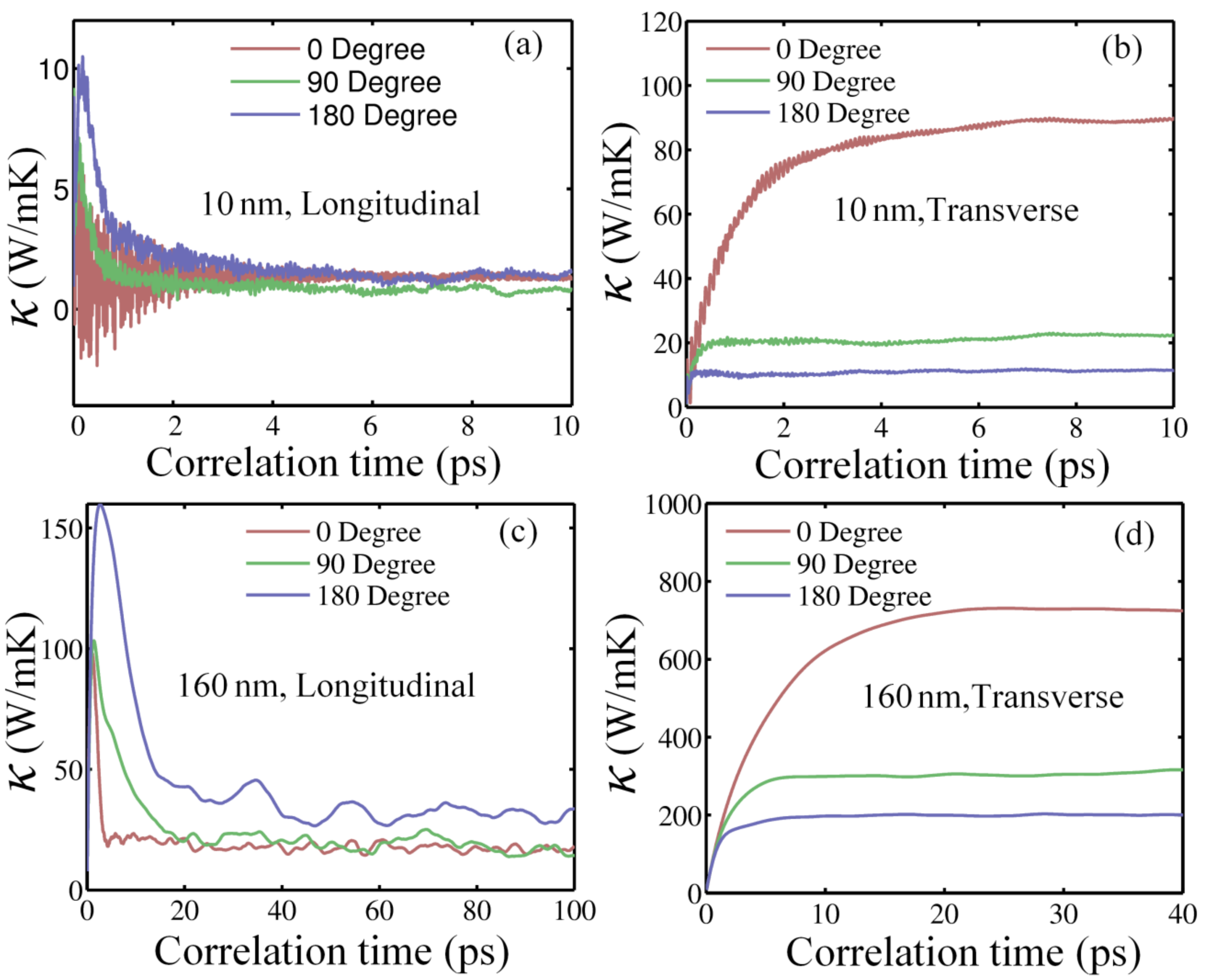}
\caption{Thermal conductivities 
of graphene kirigami at room temperature 
along the longitudinal (a,c) and transverse (b,d) directions 
as a function of correlation time, 
for samples with cut lengths of 10 nm (a,b) and 160 nm (c,d)
and curvature angles of 0, 90 and 180 degrees.
}
\label{fig2}
\end{figure}
In Fig.\ref{fig2}, typical EMD results for the calculated thermal conductivity 
along the longitudinal and transverse directions are presented as a function of correlation time, 
for graphene kirigami samples with cut lengths of 10 nm and 160 nm 
at room temperature. In this case, the volume fraction of the cuts is 20\% 
and samples with curvature angles of 
0$^\circ$, 90$^\circ$ and 180$^\circ$ for the cuts are considered. 
For samples with larger cut length, 
the predicted thermal conductivity values are considerably higher (see Fig.\ref{fig2}). 
As expected, along the longitudinal direction, since 
the cut sections interfere directly with the phonon transport, 
the thermal conductivity is much more suppressed as compared with the
transverse direction in which the cuts are basically parallel 
to the heat transfer direction. 
Moreover, based on our results along the longitudinal direction 
for sample with small cut lengths, the effects of cuts curvature angle 
on the thermal conductivity are small. 
The running thermal conductivity in this direction also exhibits 
a local peak at small correlation time, 
which is a sign of pseudo diffusive transport and localization 
caused by the combined honeycomb and kirigami lattices, 
as in the case of electron transport (see the method section). 
For the heat transfer along the transverse direction, 
by increasing the curvature angle the thermal conductivity 
decreases significantly,
which can be  related to the increased obstruction 
in the direction of the phonon transport.\\

In a recent investigation \cite{Fan_arXiv2_2016}, 
a decomposition of the thermal conductivity of
2D materials into contributions from in-plane and out-of-plane phonons was introduced. 
For pristine graphene the out-of-plane and in-plane components of the thermal conductivity 
were predicted to be ~2050 W/mK and ~850 W/mK, respectively. 
\begin{figure}[ht]
\includegraphics[width=\columnwidth]{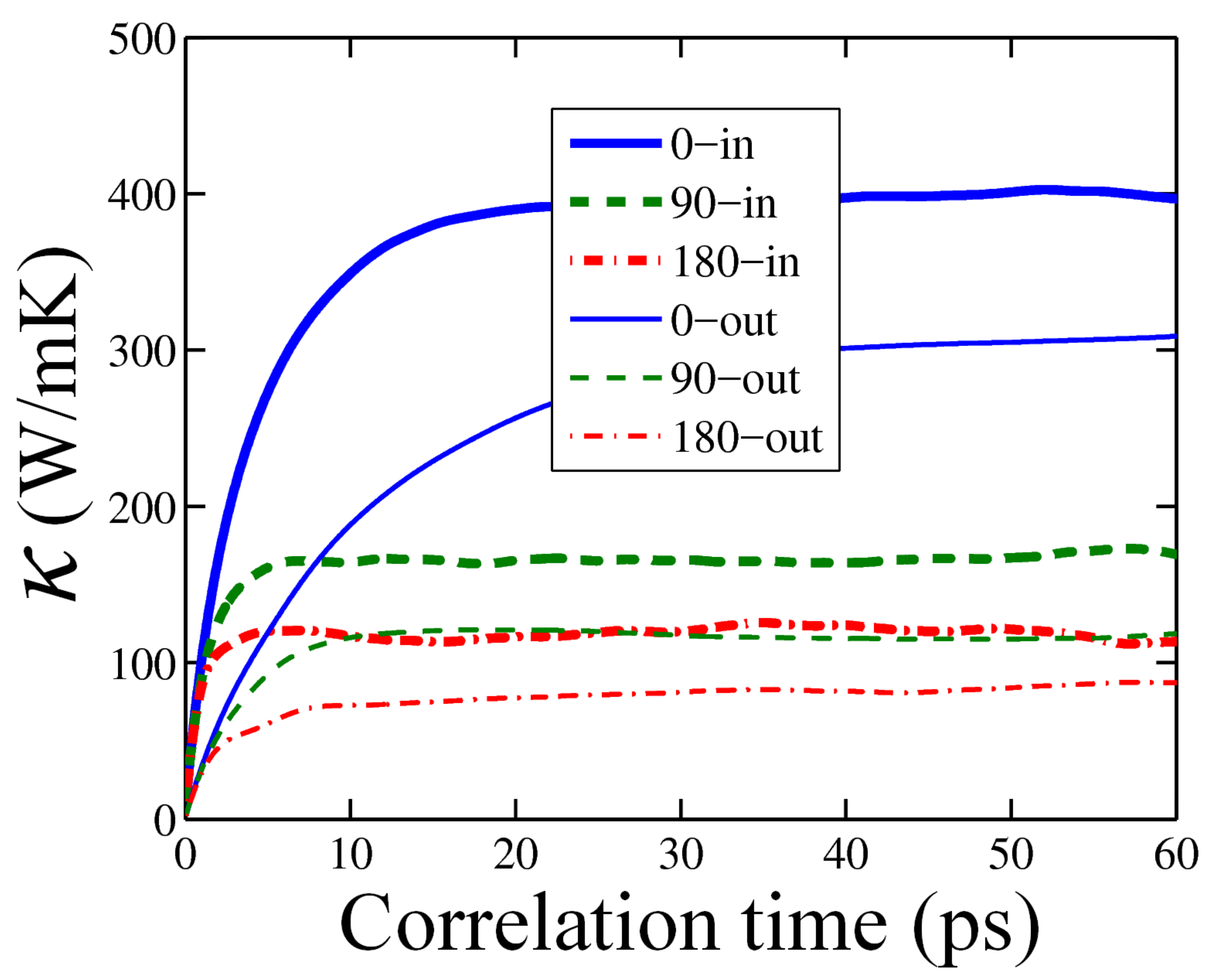}
\caption{Thermal conductivity components contributed by out-of-plane (out) 
and in-plane (in) phonons in the transverse direction for 
graphene kirigami structures with 20\% concentration of 160 nm long cuts.
}
\label{fig3}
\end{figure}
To gain more insight concerning the underlying mechanisms responsible for 
the suppression of thermal conductivity along the graphene kirigami structures, 
the calculated out-of-plane and in-plane thermal conductivity components
are compared in Fig.\ref{fig3} along the transverse direction 
for graphene kirigami networks with 20\% concentration of 160 nm long cuts. 
In contrast to the thermal transport 
in the pristine graphene, the in-plane phonons contribute more to the thermal conductivity
than the out-of-plane phonons in graphene kirigami structures. 
For example, in the case of 0 degree cuts, 
the in-plane and out-of-plane thermal conductivity components 
are about 400 W/mK and 300 W/mK, respectively, 
which are about 1/2 and 1/7 of the corresponding values in pristine graphene. 
This clearly demonstrates that the suppression of out-of-plane phonons plays 
the major role in the decline of the thermal transport along graphene kirigami films.\\

For the kirigami systems at atomic scale, 
the thermal transport is complicated since the phonons-cut scattering effects 
play the major role in the heat transport. 
However, by increasing the system size and approaching the mean-free-path of graphene 
the effect of phonon-cut scattering decreases. 
Consequently, for the kirigami structures with the large cuts in microscale 
the effective thermal conductivity approaches 
the diffusive heat transfer which can be studied 
on the basis of a continuum system that neglects 
the phonon scattering rates. 
Nonetheless, from continuum point of view, 
the effective thermal conductivity along kirigami structures 
is not trivial and to the best of our knowledge 
no analytical function exists to describe such a diffusive process.
We therefore employed finite element (FE) modeling 
to evaluate the effective thermal conductivity 
of kirigami films at the diffusive limit. 
In this study, the FE simulations were carried out using the ABAQUS/Standard package. 
As for the loading condition, we applied positive (inward) 
and negative (outward) heat-fluxes on the two opposite surfaces 
along the direction in order to investigate 
the thermal conductivity. Since the thermal transport is investigated 
along periodic structures, on the boundary edges perpendicular 
of the heat-transfer direction the temperatures of every two opposite nodes 
should be equal which can be achieved by applying 
periodic boundary condition in the FE modeling. 
We calculated the temperature gradient
along the heat transfer direction which was then accordingly 
used to compute the effective thermal conductivity 
on the basis of Fourier's law \cite{Mortazavi_Nanoscale_2014}. 
To count for atomistic effects which are related to the phonon-cut scattering rates, 
one can assume these cuts as resistors that scatter the phonons. 
In such a way, the effective thermal conductivity of 
kirigami films can be approximated by considering 
a series of line conductors (representing the graphene pristine lattices) 
that are connected by thermal resistors (representing the cuts). 
As discussed in an earlier study for the modeling of heat transfer 
in polycrystalline graphene \cite{Mortazavi_Nanoscale_2014}, 
the effective thermal conductivity of such a system 
can be represented based on the first order rational curve. 
In order to simulate the effective thermal conductivity of 
graphene kirigami structures with larger cut lengths, 
we therefore extrapolated the EMD results 
for small cut lengths using the first order rational curve. 
In these cases, we considered that the thermal conductivity 
for the kirigami films with infinite cut lengths 
is equal to that based on our FE diffusive modeling.
\begin{figure}[ht]
\includegraphics[width=\columnwidth]{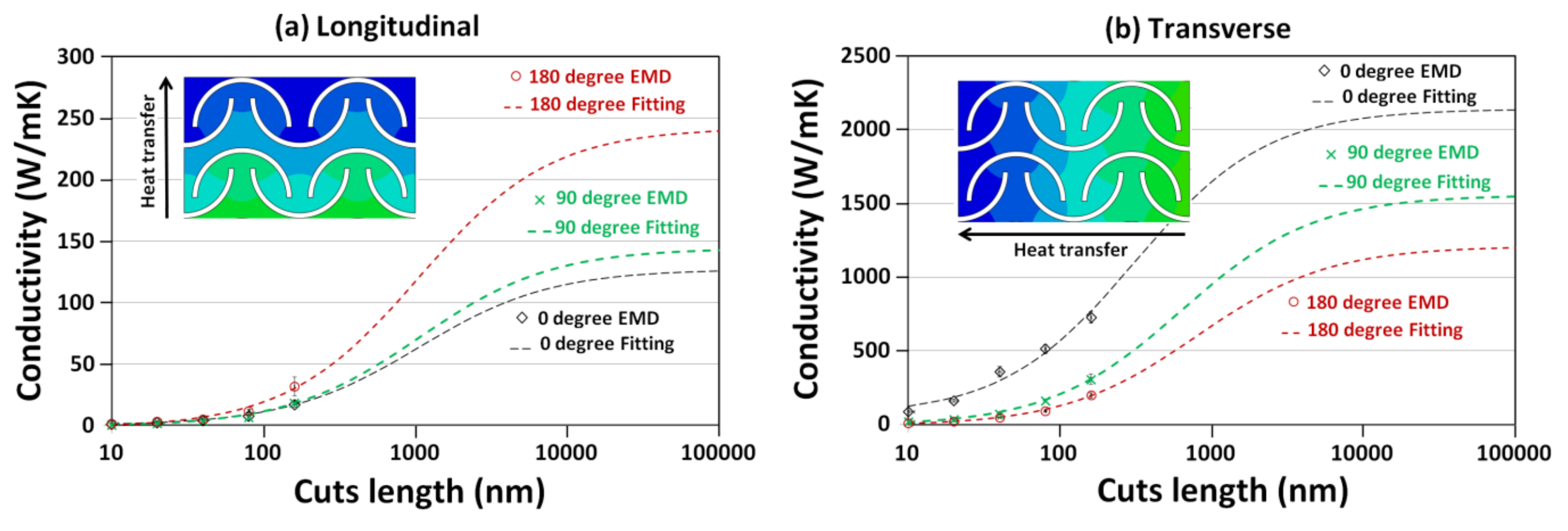}
\caption{Effective thermal conductivities of 
graphene kirigami structures at the room temperature 
as a function of cuts length along (a) longitudinal 
and (b) transverse directions. 
The points and dashed lines represent the EMD results 
and extrapolation curves, respectively. 
The insets illustrate the finite element modeling results 
for temperature distributions of systems in the diffusive regime.
}
\label{fig4}
\end{figure}
In Fig.\ref{fig4}, the calculated effective thermal conductivities of 
graphene kirigami structures by the EMD method as a function of 
cuts length along the longitudinal and transverse directions are illustrated. 
Interestingly, the devised extrapolation technique can fairly well 
reproduce the trends acquired by the fully atomistic EMD modeling. 
This result suggests that combination of 
EMD and FE modeling can be used as an efficient approach to 
estimate and engineer the effective thermal conductivity 
of graphene kirigami structures for different cut lengths 
and curvature angles as well. 
We should also note that 
the thermal conductivity of graphene can be engineered by 
at least three orders of magnitude based on the kirigami approach. 
We remind that the thermal conductivity of highly defective graphene 
so called amorphized graphene was found to be around 
two orders of magnitude smaller than that of pristine graphene \cite{Mortazavi_Carbon_2016}. 
Moreover, chemical doping was predicted to decline 
the thermal conductivity of graphene by around 
an order of magnitude \cite{Mortazavi_SSC_2012}. 
This short comparison clearly highlights that the fabrication
of graphene kirigami networks can be considered as 
a very effective method to tune the thermal transport properties
in graphene-based nanostructures.

\subsection {Stretchability}
\begin{figure}[ht]
\includegraphics[width=\columnwidth]{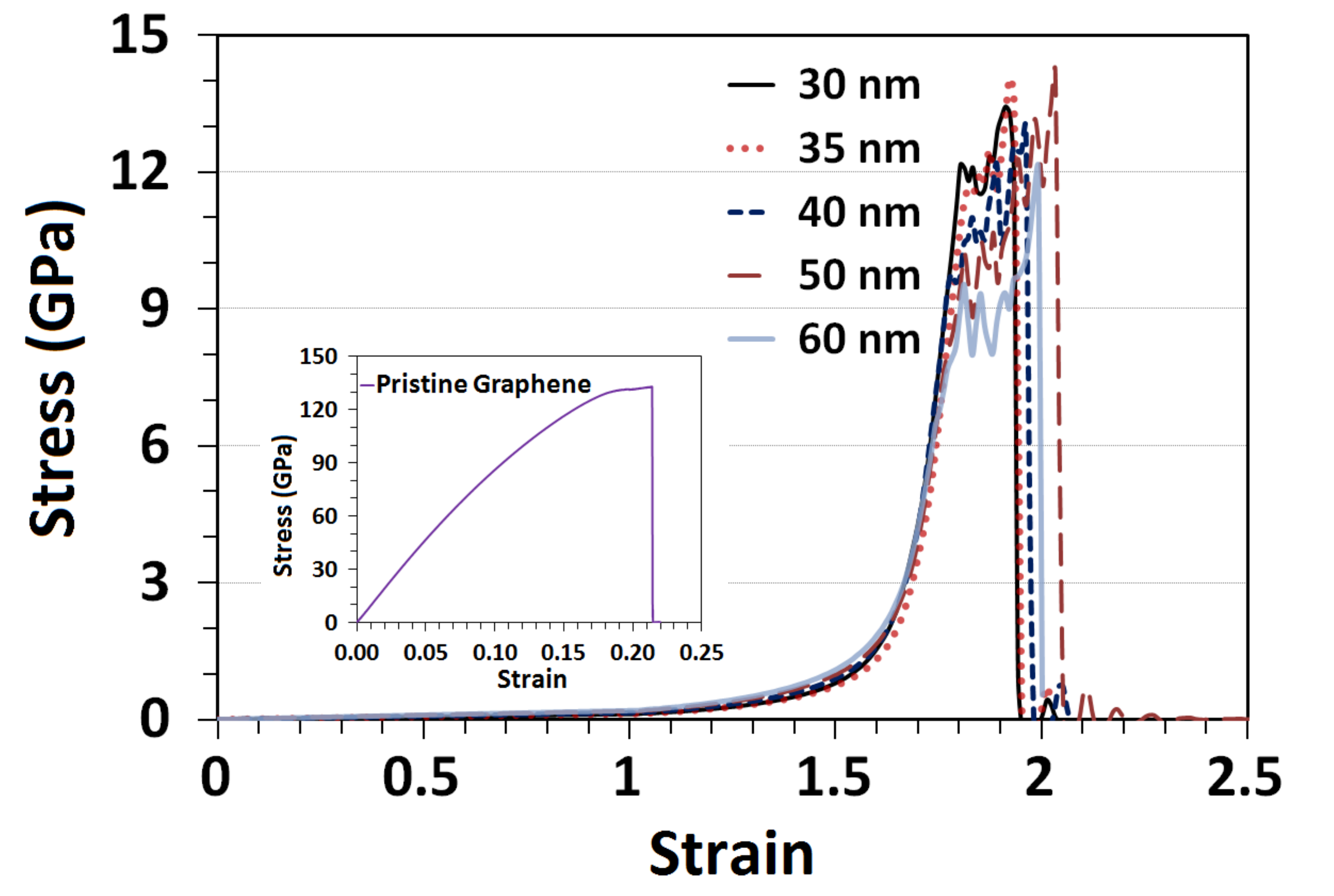}
\caption{Stress-strain responses of 
the graphene kirigami structures 
with linear cuts and with different lengths (ranging from 30 nm -- 60 nm). 
The inset illustrates the stress-strain relation of pristine graphene.
}
\label{fig5}
\end{figure}
We next shift our attention to explore 
the mechanical response and stretchability of the graphene kirigami structures. 
As observed for thermal transport, 
due to the phonon-cut scattering effects the cut length plays the major role. 
In contrast, for the stretchability of the graphene kirigami, 
the deflection of the sheets and stress-concentration 
originated from the cuts are the main factor that should be taken into account. 
To investigate the effect of cut length on the stress-strain response, 
we considered different structures with linear cuts and with different cut lengths 
and the acquired results are illustrated in Fig.\ref{fig5}.
As compared with pristine graphene (Fig.\ref{fig5} inset), 
the stress-strain responses of graphene kirigami present different patterns. 
In pristine graphene, the applied strain directly results in 
increasing of the carbon atoms bond length and therefore 
the stress value increases considerably by increasing the strain. 
On the contrary for the considered graphene kirigami structures, 
up to high strain levels of $\sim$1, the stress value only slightly increases 
and remains very low which indicates that the structure elongates 
by deflection rather than the stretching of the carbon bonds. 
After the strain level of $\sim$1, the stress values start 
to increase but smoothly which implies that the deformation 
is yet achieved by the deflection but bond elongation is also happening. 
For the considered structures, at strain levels higher than $\sim$1.6 
the stress values sharply increase which reveals that the deflection 
of the sheets contribute less and the stretching is achieved more by the bond elongation. 
Because of the limited stretchability of the carbon atoms bond, 
this step is limited and shortly after the structure reaches its failure point. 
At this stage, the stress concentration existing around the cut corners 
result in the local and instantaneous bond breakages 
thus causing instabilities and accordingly variations in the stress values. 
Around the cut corners, the bond breakages extend gradually 
and the rapture finally occurs when the two parts of the kirigami 
become completely detached by the crack coalescence. 
Since we considered the thermal effects 
by conducting the simulation at room temperature, 
the bond breakages present stochastic nature 
and therefore a sample present different stress-strain relations 
around the failure point by performing independent simulations 
with different initial velocities. 
In accordance with earlier observations ~\cite{Qi_PRB_2014}, 
our results depicted in Fig.\ref{fig5} also confirm that the stretchability 
of the graphene kirigami is convincingly independent of the cut length. 
This in an important finding and illustrates that the classical MD simulations 
can be considered as a promising approach for 
the design of graphene kirigami structures with optimized performance. 
Worthy to note that in this case, the finite element modeling 
are highly complicated mainly because of 
the existing severe out-of plane deflections 
and also stress-concentrations (that may cause singularity in the stress field). 
As it is clear for different samples, the variations for the maximum tensile 
strength is more considerable than the rupture strain of the graphene kirigami films. 
From an engineering point of view, the main goal to fabricate 
the graphene kirigami is to enhance the stretchability 
such that the values for the tensile strength are not critical. 
For the rest of our investigation, we considered the cut length of 40 nm 
and for each case we performed 4 independent simulations 
to provide the error-bars in our results.  
\begin{figure}[ht]
\includegraphics[width=\columnwidth]{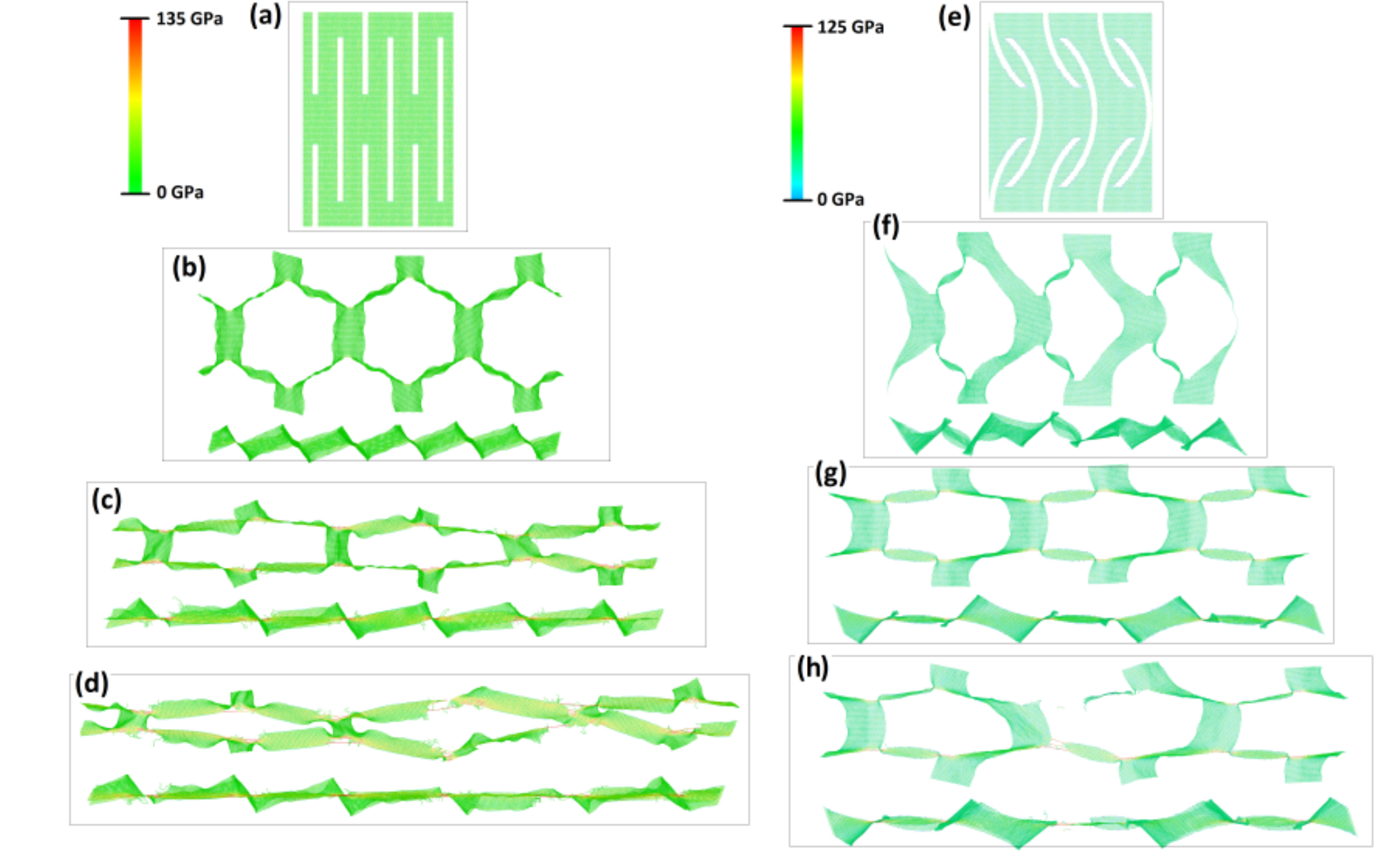}
\caption{Deformation process and 
stress distribution (in GPa) of graphene kirigami 
with periodic (a-d) linear and (e-h) 90 degree cuts. 
The OVITO package ~\cite{Stukowski_MSMSE_2010} has been used 
to illustrate these deformed structures.
}
\label{fig6}
\end{figure}
In Fig.\ref{fig6}, the deformation process of 
two different graphene kirigami structures are depicted. 
The deformation is highly symmetrical 
which verifies the accuracy of our applied loading condition 
and also confirms that the our modeling results 
are representative of graphene kirigami structures with periodic pattern of cuts. 
Moreover, up to high strain levels the stretching is achieved 
by the sheets out-of plane deflections and transverse shrinkage and 
the stress values throughout the sample are negligible, 
though in the cuts corner due to stress concentration 
they may instantaneously reach high values
(Fig.\ref{fig6}(b) and Fig.\ref{fig6}(f)). 
At higher strain levels around the failure, 
stress concentration increases around the cuts corner 
which initiate the spontaneous bond breakage (Fig.\ref{fig6}(b) and Fig.\ref{fig6}(f)), 
finally leading to the sample rupture (Fig.\ref{fig6}(d) and Fig.\ref{fig6}(h)). 
As it is clear, to design highly stretchable graphene kirigami, 
the solution is to enhance the deflection limit of the structures. 
In this regard, linear cuts are more favorable since 
they can deflect more than curved cuts. 
In addition, simple comparison based on Fig.\ref{fig6} results 
reveals that upon the stretching, the kirigami structure 
with linear cuts shrink more along the transverse direction 
in comparison with the curved cuts. 
To provide a general viewpoint concerning the engineering of 
the graphene kirigami mechanical response, 
we studied the effects of various design parameters. 
To this aim, we compared the rupture strain and 
tensile strength for different kirigami structures. 
The rupture strain is the key parameter 
which is representative of the stretchability of the graphene nanomembranes. 
The stress field in linear elastic fracture mechanics (LEFM) 
in a two dimensional solid for a straight crack can be described 
in terms of the mode I and II stress intensity factors. 
We should however emphasis that although our study is categorized 
as a large deformation problem, but the insight provided 
by the LEFM for small strain theory can be helpful 
to understand the underlying mechanism.
The stress intensity factor for the arrangement of cracks 
in transverse direction can be computed using 
the standard fracture mechanics techniques, as follows:
\begin{equation}
K_{I}=\sigma_{\infty}\sqrt{(2h)\tan\left(\frac{\pi a}{2h}\right)}
\end{equation}
Here $a$ is the cut length and $h$ is the sum of cut length 
and transverse spacing distance.
\begin{figure}[ht]
\includegraphics[width=\columnwidth]{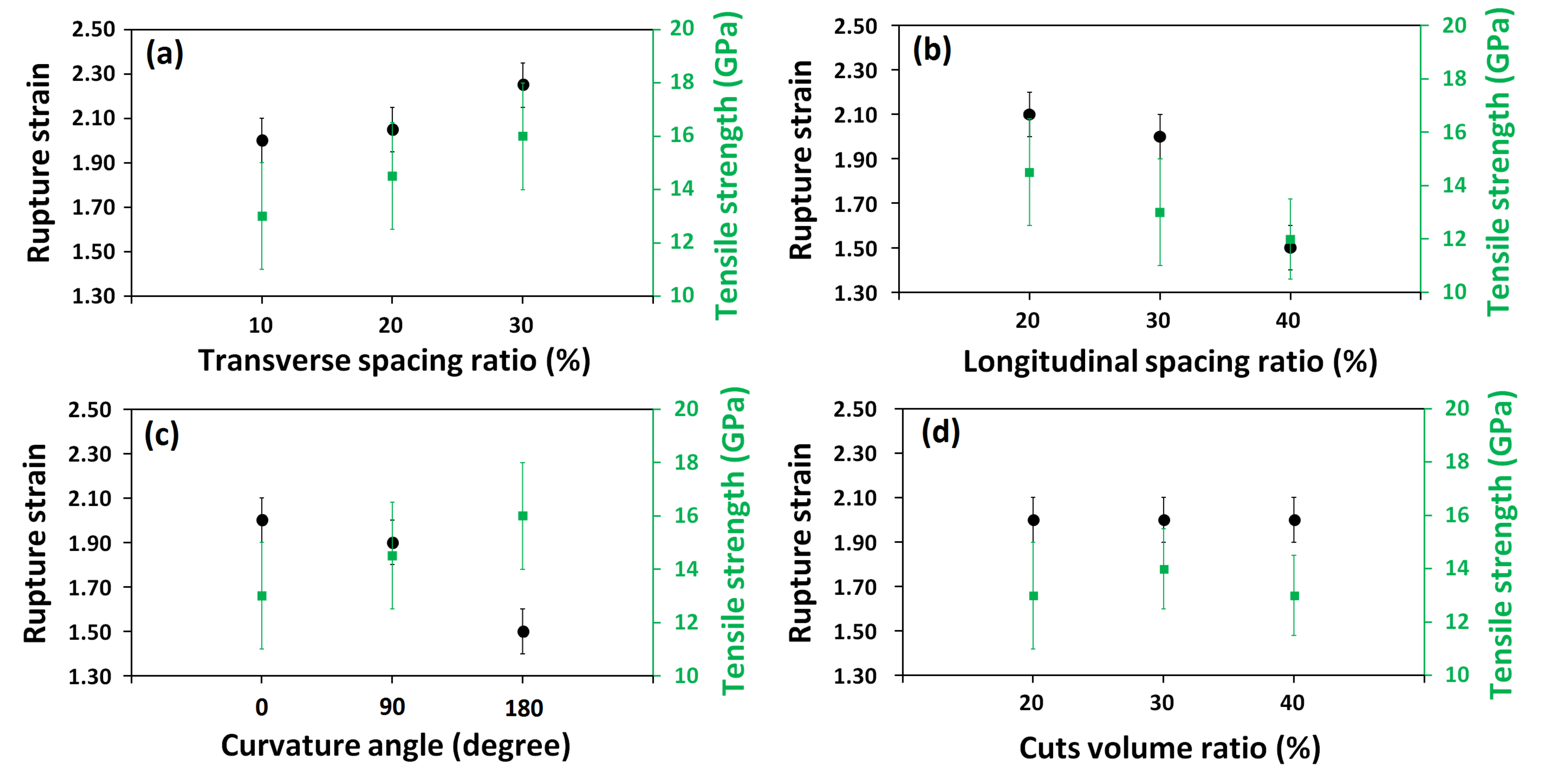}
\caption{Effect of various design parameters 
such as (a) transverse and (b) longitudinal spacing ratio,
(c) curvature angle and (d) cuts volume	ratio
on the rupture strain (stretchability) 
and tensile strength of graphene kirigami structures.
}
\label{fig7}
\end{figure}
As illustrated in Fig.\ref{fig7}(a), by increasing the transverse spacing ratio, 
both the rupture strain and the tensile strength increase. 
The increase in the maximum tensile strength is 
more considerable than the increase in the stretchability. 
The value of the stress near the cut tip decreases 
with increasing the transverse distance. 
Therefore, the larger the transverse spacing between the cracks, 
the less is the influence from the interaction of 
the near tip stress fields. According to Eq. 3, 
the stress-intensity factor decreases by increasing 
the transverse spacing distance due to suppression of 
the interaction of the stress fields of adjacent cuts. 
Consequently, by increasing the transverse spacing ratio 
the stress-concentration decreases and such that 
the tensile strength increases. 
Note that for a very high transverse spacing 
the graphene kirigami approaches to the graphene structure 
with few cracks inside, which yields to a failure strain lower 
than that of the pristine graphene. 
For the graphene kirigami accordingly there should be 
an optimum transverse spacing distance that 
maximizes the stretchability. 
The situation is opposite for the increase of 
longitudinal spacing ratio. 
Indeed, by increasing the longitudinal spacing the stretchability drops significantly. 
This is an expected trend because the stretchability of the kirigami structures 
is basically originated from the longitudinal spacing. 
By increasing the longitudinal spacing, 
the out-of-plane deflection decreases and 
the structure approaches to the graphene film with cracks. 
Regarding the cuts curvature angle, 
we found that the linear cuts present the highest stretchability 
as illustrated in Fig.\ref{fig7}(c). For the linear cuts, 
the rupture strain is higher than the one 
for curved cracks due to the fact that the effective distance 
between two curved cuts becomes much smaller. 
Hence, for the curved cuts the film deflection decreases 
and additionally the crack path until coalescence is shorter 
and therefore the stretchability is not as high as for the linear cuts. 
Nonetheless, it has been shown by Karihaloo ~\cite{Karihaloo_MechMat_1982}, 
that the stress intensity factor for a curved crack compared to 
that of the straight crack of same length being positioned 
at the same location and subjected to far field stress is lower. 
Therefore by increasing the cuts curvature angle, 
the stress concentration decreases and such that the tensile strength increases. 
Interestingly, for the cuts volume fraction 
we found that it does not yield considerable effects 
neither on the stretchability nor on the tensile strength (see Fig.\ref{fig7}(d)). 
Based on our brief parametric investigation, 
the transverse and longitudinal spacing ratios can be considered 
as the two main factors that could be optimized to increase both, 
stretchability and tensile strength. 
Our results also clearly confirm that the graphene kirigami structures 
can be stretchable by more than 10 times as compared with the pristine graphene.

\subsection {Electronic structures}\label{section_electronic_structure}

\begin{figure}[ht]
\includegraphics[width=\columnwidth]{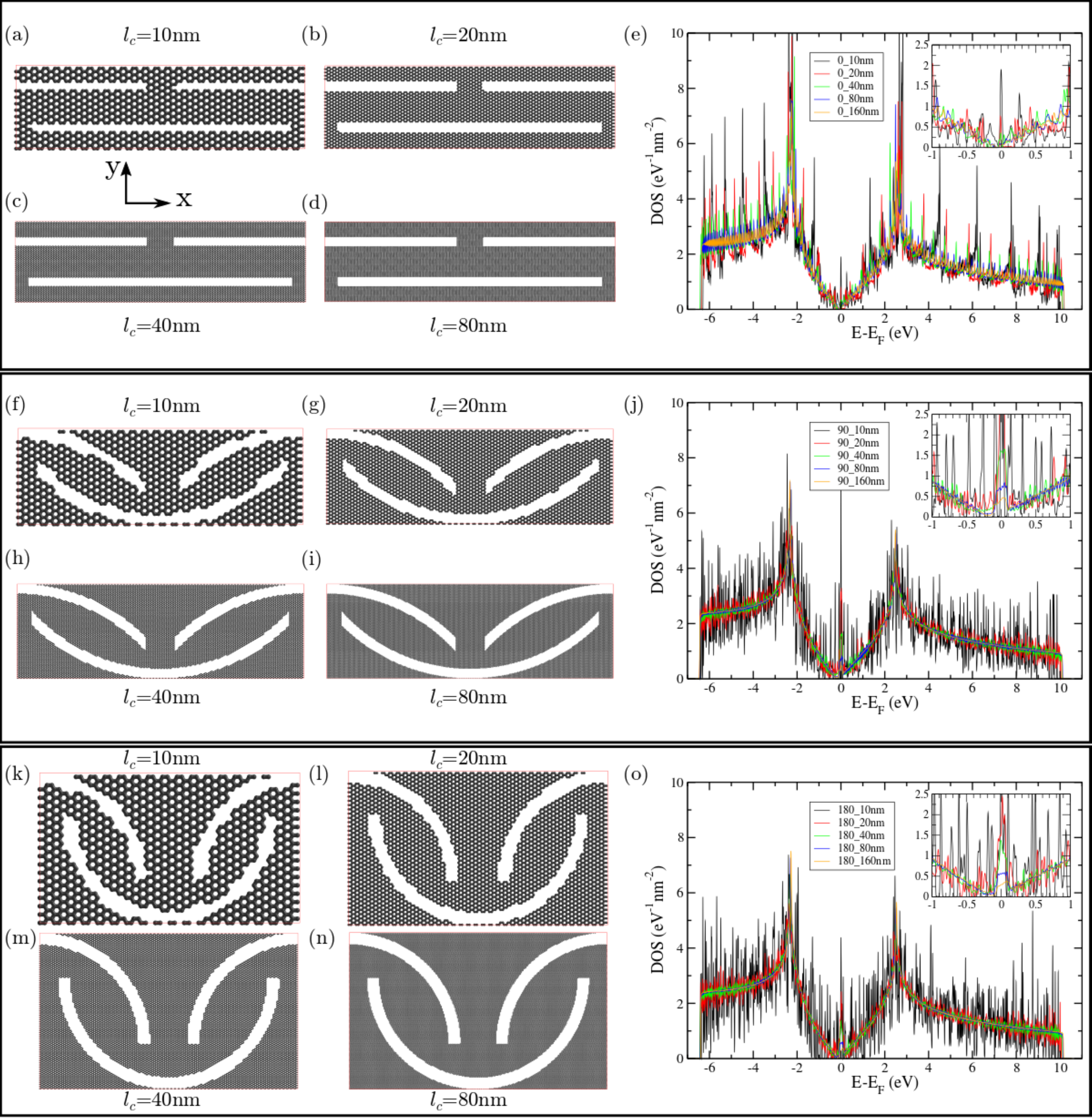}
\caption{Atomic structures of graphene kirigami obtained with various
cut angles ($\theta_c$) and cut lengths ($l_c$) and their associated DOS. 
Top panels (a-e): Unit cells of graphene kirigami corresponding to $\theta_c$=0$^\circ$ and
$l_c$=10, 20, 40, and 80 nm (a),(b),(c), and (d), respectively and the associated DOS (e).
Middle panels (f-j): Unit cells of graphene kirigami corresponding to $\theta_c$=90$^\circ$ and
$l_c$=10, 20, 40, and 80 nm (f),(g),(h), and (i), respectively and the associated DOS (j).
Bottom panels (k-o): Unit cells of graphene kirigami corresponding to $\theta_c$=180$^\circ$ and
$l_c$=10, 20, 40, and 80 nm (k),(l),(m), and (n), respectively and the associated DOS (o).
}
\label{fig8}
\end{figure}

After having explored their thermal and mechanical properties,
we now focus on the electronic structure properties of the graphene kirigami.
First, the density of states (DOS) of various graphene
kirigami structures are presented in Fig.\ref{fig8}.
The total widths/lengths of the large systems 
vary between 320 and 390 nm and are obtained by repeating the unit cells 
as illustrated in Figs.\ref{fig8}(left panels).
The total size of the system is large enough to ensure
an accurate energy resolution within the Lanczos technique
and also a rapid convergence of the average on random phase wave packets.
Note that three different angles were considered for the cuts, 
$\theta_c$=0, 90, and 180$^{\circ}$.
For $\theta_c$=0$^{\circ}$, the cuts are actually straight (Fig.\ref{fig8}(a-d)),
while for 90 and 180$^{\circ}$, the cuts are curved (Fig.\ref{fig8}(f-i,k-n)).
For each $\theta_c$, five different cut lengths are considered,
$l_c$=10, 20, 40, 80, and 160 nm (the unit cells of the $l_c$=160 nm systems
are not shown in Fig.\ref{fig8}).

The DOS of the case $\theta_c$=0$^{\circ}$, for different cut lengths
$l_c$ are presented in Fig.\ref{fig8}(e), while the DOS corresponding to 
$\theta_c$=90 and 180$^{\circ}$ are depicted 
in Fig.\ref{fig8}(j) and Fig.\ref{fig8}(o), respectively.
One recognizes easily the global shape of the graphene DOS 
corresponding to the $\pi$ and $\pi^{*}$ bands, 
i.e. a Heaviside function at the band edges ($\Gamma$ point in graphene Brillouin zone (BZ)),
two van Hove singularities ($M$ point in the BZ), and
a V-shape, i.e. a linear increase as a function of energy 
around the Dirac point ($K$ point in the BZ).
However, one also obviously notices the presence of many DOS sub-peaks
which can be attributed to van Hove singularities typical of 1D systems,
which appear because of the formation of \textit{connected} nanoribbons 
generated by the cuts in graphene.
The energy spacing between these 1D van Hove singularities is 
inversely proportional to the ribbon's width.
For very thin ribbons ($l_c=$10 nm cases, see for instance  Fig.\ref{fig8}(a)),
the DOS is quite spiky and deviates considerably from the pristine graphene DOS,
in particular close to the Fermi energy (Dirac point in graphene)
where the DOS sub-structure is complex (see insets in Fig.\ref{fig8}(e,j,o)).
In particular, small energy gaps can appear as observed for instance
for $(\theta_c,l_c)=$ (90$^\circ$,10 nm) and (180$^\circ$,10 nm) while a more significant
gap is observed for $(\theta_c,l_c)=$ (0$^\circ$,20 nm) in the energy range [-0.2,+0.2] eV
with a peak centered on the zero energy.
However, when increasing $l_c$ the DOS recovers progressively the
shape of pristine graphene DOS although residual oscillations 
coming from the 1D van Hove singularities are still present.
Note that for $\theta_c$=90 and 180$^{\circ}$, although the DOS 
become similar to graphene DOS, a non-negligible peak remains
present around the Fermi energy (see for instance $l_c$=80, 160 nm curves 
in the insets of Fig.\ref{fig8}(j,o)).
Those persistent peaks are originating from strongly localized states,
most probably located close to the cut edges,
similarly to what was observed for vacancies in graphene~\cite{Cresti_Crystals_2013}.
Such localized states can affect the electronic transport
and can possibly induce a transport gap in the corresponding energy window.

\subsection {Electronic transport}\label{section_electronic_transport}

Recently, Bahamon \textit{et al.} discussed the electronic transport 
in 34 nm long and 10 nm wide 1D graphene kirigami structures
with straight cuts using the Landauer-B\"{u}ttiker formalism~\cite{Bahamon_PRB_2016}.
They observed that at this nanoscale size the 1D electronic transport
is governed by resonant tunneling through coupled localized states acting as quantum dots.
Upon moderate elongation (15\%), the conductance profile is first degraded 
because of the decoupling of the states
but it can revive at a larger elongation (30\%).
\begin{figure}[ht]
\includegraphics[height=0.92\textheight]{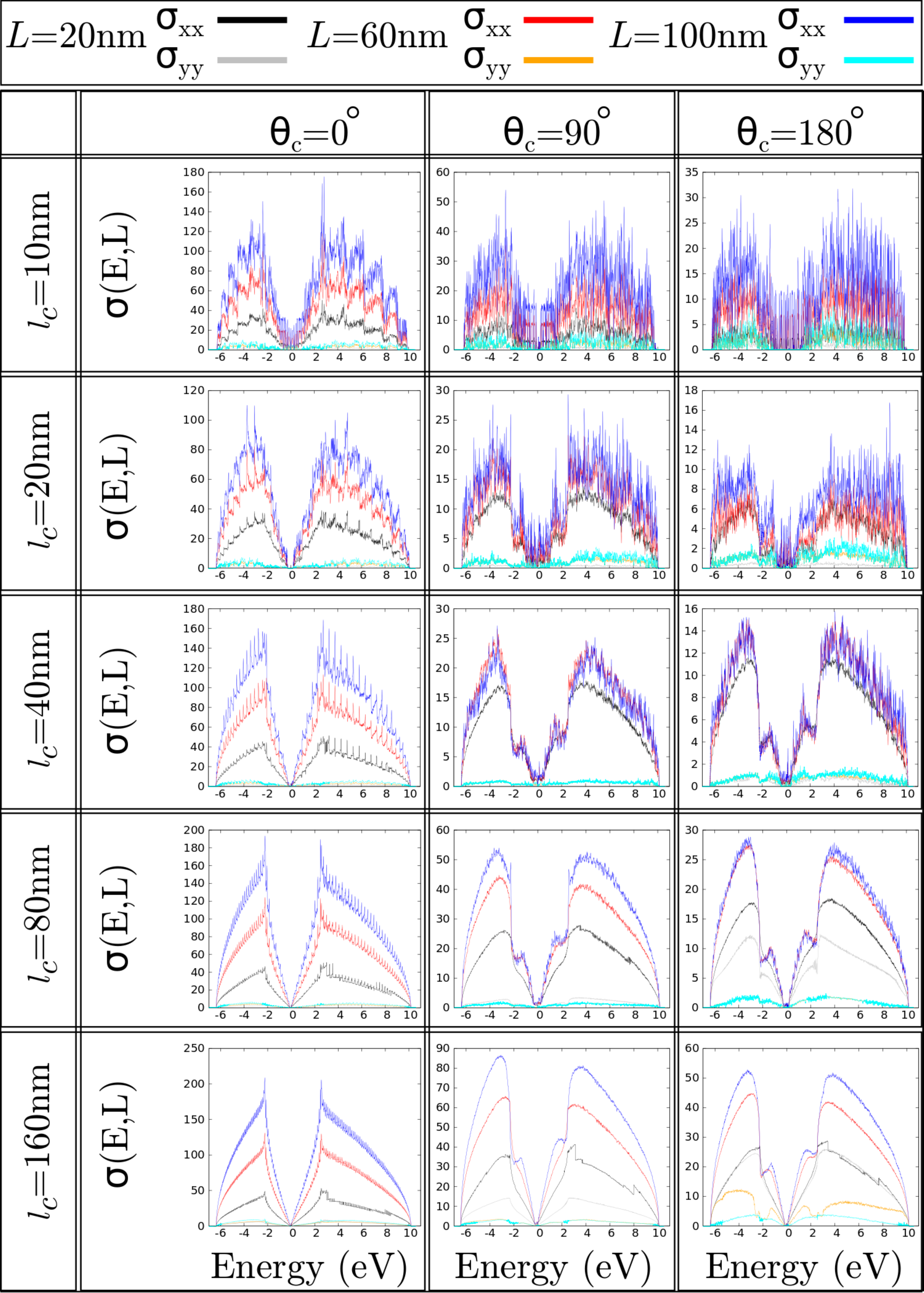}
\caption{Electronic conductivity ($\sigma(E,L)$) 
in units of $G_0=2e^2/h$ for various kirigami structures with
(from left to right) cut angles $\theta_c$=0, 90, and 180$^{\circ}$,
(from top to bottom) cut lengths $l_c$=10, 20, 40, 80, and 160 nm.
}
\label{fig9}
\end{figure}
Here, the electronic transport properties of unstretched 2D periodic graphene 
kirigami are investigated using the real-space Kubo-Greenwood method.
As discussed in the method section, the Kubo-Greenwood formalism allow us to describe 
the electronic diffusion and the associated conductivity.
The diffusivity curves are presented in the supplementary information document.
They illustrate the discussion conducted in the method section regarding
the presence of the two-scale periodicity and its impact 
on the ballistic and intermediate diffusive/localized transport regimes.
However, in order to gain more insight and obtain 
a more concrete picture of the electronic transport properties
in these systems it is preferable to characterize them through the energy- and length-dependent
Kubo-Greenwood conductivity $\sigma(E,L)$.
The longitudinal components of the conductivity tensor
$\sigma_{\text{xx}}(E,L)$ and $\sigma_{\text{yy}}(E,L)$ have been calculated
for three specific propagation length $L=$ 20, 60 and 100 nm and are presented in Fig.\ref{fig9}.
One can immediately notice three points. First, it is clear that 
the conductivities along the Y direction (gray, orange and cyan curves) 
are much lower than the conductivities along the X direction (black, red and blue curves).
Second, the increase of the cut angle from $\theta_c$=0$^\circ$ to $\theta_c$=180$^\circ$
(i.e. from left to right panels) decreases the overall conductivity profile in absolute values.
Finally, the transition from a 1D nanoribbon-like to a 2D graphene-like 
electronic transport is clearly visible when going 
from small to large cut lengths $l_c$ (i.e. from top to bottom panels)
as the 1D subbands structure is progressively smoothened out.
The conductivities $\sigma_{\text{xx}}(E,L)$ and $\sigma_{\text{yy}}(E,L)$ 
are found to globally increase with propagation length 
although the increase is not really significant for $\sigma_{\text{yy}}(E,L)$ which is maintained to low values.
Actually for high cut angle and long cut lengths, $\sigma_{\text{yy}}(E,L)$ displays a decreasing behavior
synonym of localization effects.
Note also that for $\sigma_{\text{yy}}$ the propagation length $L=$100 nm is not reached at all energies $E$
because of the slow propagation velocities in this direction.
As for thermal transport, these results demonstrate a strong anisotropy 
between the two principal electronic transport directions.
Then, as noticed in the DOS, the presence of a band gap is clearly observed in the system $(\theta_c,l_c)$=(0$^\circ$,20 nm).
Interestingly, the general shape of $\sigma_{\text{xx}}$ for curved cuts
is different from the $\theta_c$=0$^\circ$ case. Indeed,
the electron/hole regions (positive/negative charge carrier energies) exhibit two main peaks
instead of only one. There is no particular features in the DOS 
that would help to explain this double peak structure.
Actually, it is interesting to note that for $(\theta_c,l_c)$=(180$^\circ$,160 nm) 
the conductivity curve exhibit a single peak structure 
at small propagation length ($L=$20 nm) before showing a double peak structure
at longer propagation length ($L=$100 nm). 
Overall, a conductivity ratio $\frac{\sigma_{\text{xx}}}{\sigma_{\text{yy}}} \sim$ 10--20 can be estimated.

\section{Concluding remarks}
Extensive atomistic simulations were conducted to explore 
the electronic and thermal transport as well as the stretchability of graphene kirigami structures. 
The properties of graphene kirigami structures were investigated using 
periodic patterns of linear and curved cuts. 
Equilibrium molecular dynamics (EMD) method was used
to calculate the effective thermal conductivity of 
graphene kirigami structures at atomic scale. 
The thermal transport is found to be highly anisotropic 
and the phonons-cut scattering effect plays the major role in the heat transport. 
The effective thermal conductivity of graphene kirigami structures 
as a function of cut lengths was established by extrapolating the EMD results 
for small cut lengths to the estimation based on the finite element (FE) diffusive modeling. 
It was confirmed that the proposed combination of EMD and FE modeling 
can be accurately used to evaluate the effective thermal conductivity 
of graphene kirigami structures from nano to macro scale. 
Notably, our results suggest that the thermal conductivity of 
graphene can be engineered by more than three orders of magnitude 
based on the kirigami approach. 
Then the mechanical response of graphene kirigami structures 
were investigated by performing uniaxial tensile molecular dynamics simulations. 
Up to high strain levels, the kirigami structure elongates 
by out-of-plane deflection rather than bond stretching 
and thus the stress values remain negligible. 
The stress values later start to increase smoothly implying 
that bond elongation also occurs. 
Before the failure point, the stress values increase sharply 
and the existing stress concentrations around the cuts corner 
result in bond breakages and finally leads to the sample rupture. 
Consequently, the stretchability of the graphene kirigami 
was suggested to directly correlate to the out-of-plane deflection limit 
of the structure which can be improved by using 
the straight cuts with optimized transverse and longitudinal spacing ratios.
Finally, electronic transport was investigated. 
The electronic density of states displayed series of van Hove peaks 
which are signature of one-dimensional confinement in case 
of graphene kirigami at the nanoscale. 
Band gaps can even occur in some cases, as well as transport gaps
caused by strongly localized states especially in curved cuts.
Finally, electronic conductivities are found to be highly
anisotropic with a ratio of $\sim$ 10--20. Unfortunately,
this electronic anisotropy is analogous to the
thermal anisotropy which is in principle not 
in favor of thermoelectric applications. 
However, as discussed, the tunability of graphene kirigami properties
is very large and therefore may open routes towards thermoelectric engeenering and
new kinds of ultra thin NEMS with superior properties. 

\section {Acknowledgements}
B.M. and T.R. greatly acknowledge the financial support 
by European Research Council for COMBAT project (Grant number 615132).
Z.F. and A.H. are supported by the National Natural Science Foundation of
China (Grant No.  11404033) 
and the Academy of Finland through its Centres of Excellence Program (project no. 251748) 
and they acknowledge the computational resources 
provided by Aalto Science-IT project and Finland's IT Center for Science (CSC).
A.L., and J.-C.C. acknowledge financial support 
from the F\'{e}d\'{e}ration Wallonie-Bruxelles through the ARC entitled 3D Nanoarchitecturing of 2D crystals (N$^o$ 16/21-077),
from the European Union's Horizon 2020 research and innovation programme (N$^o$ 696656),
and from the Belgium FNRS.
Computational resources have been partly provided by 
the supercomputing facilities of the Universit\'e catholique de Louvain (CISM/UCL)
and the Consortium des \'Equipements de Calcul Intensif en F\'ed\'eration Wallonie Bruxelles
(C\'ECI) funded by the Fond de la Recherche Scientifique de Belgique (F.R.S.-FNRS) under convention 2.5020.11

\bibliography{graphenekirigami}

\clearpage

\newpage
\section {Supplementary information}\label{SuppInfo}

\subsection{Electronic diffusivity}
\begin{figure}[ht]
\includegraphics[width=\columnwidth]{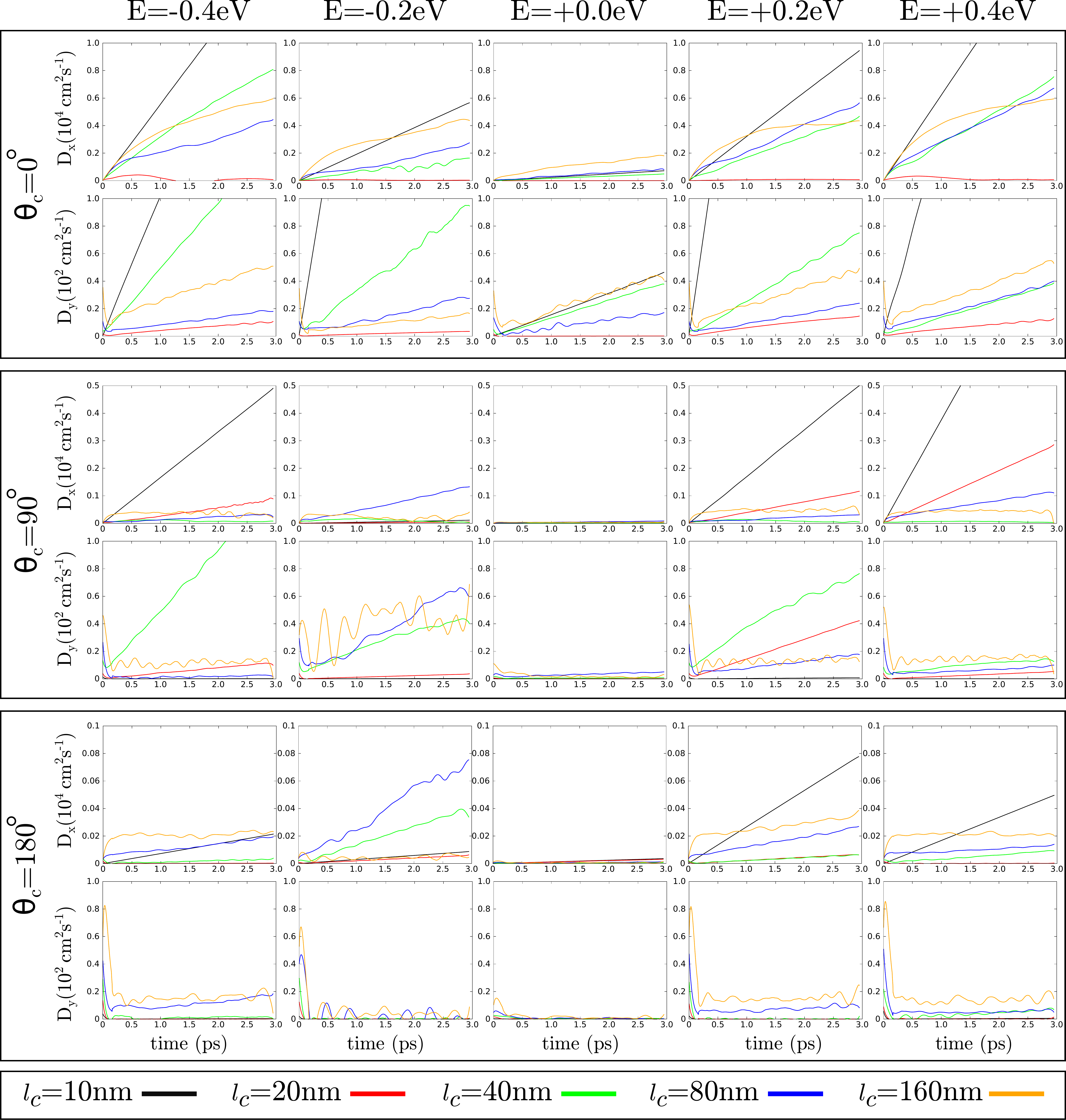}
\caption{Energy and time dependent 
diffusivity ($D_{\text{x}}(E,t)$ and $D_{\text{y}}(E,t)$) for different
kirigami structures with
cut angles $\theta_c$=0, 90, and 180$^{\circ}$,
cut lengths $l_c$=10, 20, 40, 80, and 160 nm.
}
\label{supfig1}
\end{figure}

In order to display the complex dynamics of charge carriers 
taking place in the kirigami structures one discusses here the directional, energy and time dependent diffusivities.
The diffusivities along the x direction ($D_{\text{x}}(E,t)$) and along the y direction ($D_{\text{y}}(E,t)$)
are presented in Fig.\ref{supfig1} for five selected energies 
$E=$-0.4, -0.2, 0.0, +0.2, and +0.4 eV with respect to the Fermi energy ($E_F$).
The top panels of Fig.\ref{supfig1} correspond to $\theta_c$=0$^\circ$,
the middle panels to $\theta_c$=90$^\circ$, and the bottom panels to $\theta_c$=180$^\circ$.
First, one notices that $D_{\text{y}}(E,t)$ is systematically 
lower than $D_{\text{x}}(E,t)$ by one or two orders of magnitude,
indicating a clear anisotropy in the two transport directions.
Second, the magnitude of the diffusivity is dependent on the energy which
can be understood from the complex DOS containing numerous van Hove singularities
and possible band gaps as observed in main text in Fig.\ref{fig8}.
Some diffusivity curves exhibit significant undulations which can introduced
by the numerical derivatives of wave packet quadratic spreading 
($\frac{\partial \Delta X^2(E,t)}{\partial t}$, $\frac{\partial \Delta Y^2(E,t)}{\partial t}$),
in particular for long cut lengths $l_c=$80 and 160 nm.
In some cases, it even induces a negative value of the diffusivity because 
in the localization regime the saturating quadratic spreading 
is no more increasing monotonically and may slightly oscillate, i.e. has a small negative slope.
This is a possible artifact of the calculation.
One also observes that the long time ballistic regime 
discussed in the main text in section \ref{section_methods} 
is easily observed for $l_c=$10 nm (red curves)
but is not clearly obtained for the other systems. 
For $(\theta_c,l_c)$=(0$^\circ$,20 nm), the diffusivities
are found to be anomalously low compared to other systems.
However, this can be rationalized by the fact that a rather large band gap
with localized states were observed in the vicinity of the Fermi level
for this $(\theta_c,l_c)$=(0$^\circ$,20 nm) graphene kirigami.
Some diffusivity curves display a diffusive regime i.e. a  constant value of $D(t)$
(see for instance $D_{\text{x}}(E,t)$ and $D_{\text{y}}(E,t)$ of the 
$(\theta_c,l_c)$=(180$^\circ$,160 nm) for $E$=-0.4, +0.2 and +0.4 eV),
which results from the scattering at the cut edges.
In other cases, a localized regime is observed i.e. a zero value of $D(t)$
corresponding to constructive quantum interferences in scattering loops
inducing localization of the wave packet.
However, as mentioned in main text in section \ref{section_methods}, at 
longer propagation time the diffusivity 
should finally recover a ballistic behavior 
because of the periodicity.
Overall, Fig.\ref{supfig1} demonstrates the complex transient regimes
that can be obtained in these graphene kirigami structures.\\

\end{document}